\def\fr#1#2{\hbox{${#1\over #2}$}}  
\def\subsub#1{\vskip.3cm{\bf #1}}
\def\ni{\noindent}       
\def\vs{\vskip.3cm}  
\def\Tr{\,{\rm Tr}\,}
\def\cdt{\hskip-2pt\cdot\hskip-2pt}
\def\Ra{\Rightarrow}
\def\+{ {(+)} }  \def\-{ {(-)} }   \def\0{ {(0)} }
\def\1{ {(1)} }  \def\2{ {(2)} }   
\def\pd{\partial}           \def\hpd{\hat\partial}

\def\m{\mu}              \def\n{\nu}              \def\k{\kappa}
\def\G{\Gamma}           \def\g{\gamma}           \def\d{\delta}
\def\S{\Sigma}           \def\s{\sigma}           \def\t{\tau}  
\def\a{\alpha}           \def\b{\beta}            \def\th{\theta}
\def\vphi{\varphi}       \def\Y{\Upsilon}
\def\D{\Delta}                   
\def\con{\omega}         \def\hcon{\hat\omega} 
\def\cs{\omega}
\def\ve{\varepsilon}           \def\O{\Omega}       
\def\bve{{\bar\varepsilon}}    \def\bO{\bar\Omega}
\def\bE{\bar E}          \def\bv{{\bar v}}    
\def\hth{\hat\theta}     \def\hg{\hat g}          
\def\hnabla{\hat\nabla}  \def\hpd{\hat\partial}
\documentstyle[11pt,eqs]{article}

\renewcommand{\theequation}{\thesection.\arabic{equation}}
\def\cleq{\setcounter{equation}{0}}
\def\nn{\nonumber} 
\def\be{\begin{equation}}             \def\ee{\end{equation}}
\def\ba{\begin{array}{rcl}}           \def\ea{\end{array}}
\def\beqa{\begin{eqnarray} }          \def\eeqa{\end{eqnarray} }
\def\beqalign{\begin{eqalign}}        \def\eeqalign{\end{eqalign}}
\def\leq#1{\label{eq:#1}}             \def\eq#1{(\ref{eq:#1})}
\def\bsubeq{\begin{subequations}}     \def\esubeq{\end{subequations}}
\def\bitem{\begin{itemize}}           \def\eitem{\end{itemize}}  
\begin{document}

\title{2D Induced Gravity as an Effective \\  
          WZNW System\thanks{Work supported in part by the Serbian 
                             Science Foundation, Yugoslavia} }
\author{M. Blagojevi\'c and B. Sazdovi\'c\thanks{
       e-mail addresses: mb@phy.bg.ac.yu, and sazdovi\'c@phy.bg.ac.yu}\\
       Institute of Physics, 11001 Belgrade, P.O.Box 57, Yugoslavia}       
\date{}
\maketitle
\begin{abstract} 
We introduced a dynamical system given by a difference
of two simple $SL(2,R)$ Wess--Zumino--Novikov--Witten (WZNW) actions in
2D, and defined the related gauge theory in a consistent way. It is
shown that gauge symmetry can be fixed in such a way that, after
integrating out some dynamical variables in the functional integral,
one obtains the induced gravity action.   
\end{abstract}
\vs
\ni {\it PACS number(s)\/}: 04.60 Kz, 11.10 Kk, 11.15 q \par
\ni {\it Key words\/}: two--dimensional, Wess--Zumino, gauge, 
                       induced gravity 
                                  
\section{Introduction} 

Two--dimensional gravity naturally appears in the string functional
integral in subcritical dimension, where it represents an effective
theory of quantum fluctuations of matter fields coupled to the metric
of the string world sheet \cite{1}. The induced, effective action is closely
related to the Weyl anomaly of the original string theory, and
represents a gravitational analogue of the usual Wess--Zumino action in
gauge theories. Dynamical structure of 2D gravity is, therefore, an
important aspect of string theory, but it also represents a useful
model for the theory of gravitational phenomena in four dimensions.    

Polyakov and his collaborators \cite{2} demonstrated that in the light--cone
gauge the $n$--point functions of the effective 2D gravity can be
explicitly found. Although the gauge is fixed, these solutions display
a hidden chiral $SL(2,R)$ symmetry, which turned out to be very
important for the analysis of quantum dynamics.  
These results motivated the investigation of the structure of 2D gravity
in the conformal gauge, where it becomes the standard Liouville theory
\cite{3}. Although the $SL(2,R)$ symmetry is naturally connected to the
light--cone gauge, there exists a canonical formulation of the theory
in terms of gauge independent variables, the $SL(2,R)$ currents, which
demonstrates the importance of this symmetry for the general
structure of the theory \cite{4}.

Dynamical significance of the $SL(2,R)$ symmetry, and strong analogy
between the induced gravity and the usual Wess--Zumino action, inspired
detailed investigations of the relation between the $SL(2,R)$
Wess--Zumino--Novikov--Witten (WZNW) theory and the induced gravity.
Polyakov found the connection between the $SL(2,R)$ WZNW theory and the
induced gravity in the {\it light--cone gauge\/} \cite{5}. Similar results
in the light--cone gauge have been also obtained in Refs. \cite{6,7}. The same
problem was discussed in the {\it conformal\/} gauge in Ref.\cite{8}, where
it was shown that Liouville theory may be regarded as the WZNW theory,
reduced by certain conformally invariant constraints. These constraints
can be automatically produced if one considers gauge extension of
the $SL(2,R)$ WZNW model, based on two gauge fields \cite{9}.  

In a recent letter \cite{10} we used a general method of constructing
canonical gauge invariant actions to establish the connection between
2D induced gravity and a WZNW system, defined by a difference of two
simple WZNW actions for $SL(2,R)$ group: 
\be
I(g_1,g_2)=I(g_1)-I(g_2)\, ,\qquad g_1,g_2\in SL(2,R)\, .    \leq{1.1}
\ee 
In this paper we set up the Lagrangian framework for this connection,
starting from a gauge invariant extension of the WZNW system \eq{1.1}.
The connection is established in a {\it covariant\/} way, fully
respecting the diffeomorphism invariance of both theories.  
The approach will be very useful for constructing and studying
properties of general solutions of the induced gravity, in terms of the
related simpler solutions of the WZNW system.  

After recalling some basic properties of the WZNW theory in
Section 2, we introduce in Section 3 a consistent formulation of the
gauge invariant extension of our basic object, the WZNW system \eq{1.1}.  
By taking the difference of {\it two\/} WZNW actions we are able to
overcome the usual difficulties which one encounters in the process of
gauging a {\it single\/} WZNW theory \cite{11,12}. In Section 4 we
explicitly choose the gauge group, a four--parameter subgroup of
$SL(2,R)\times SL(2,R)$ leading to four gauge fields, which is
sufficient for generating an effective transition to the induced
gravity. In Section 5 we show that new gauge invariance can be fixed in
such a way that, after integrating out some variables, one arrives at
the induced gravity action:    
\be
I(\phi,g_{\m\n})=\int d^2\xi \sqrt{-g}\,\bigl[ 
    \fr{1}{2}g^{\m\n}\pd_\m\phi\pd_\n\phi
   +\fr{1}{2}\a\phi R -M\bigl( e^{2\phi/\a} -1\bigr) \bigr]\, .\leq{1.2}
\ee
In this process, the original symmetry of the action \eq{1.1} under
conformal rescalings is also fixed. Appendices A, B and C are devoted
to some details concerning geometrical properties of spacetime $\S$ and
the group manifold $SL(2,R)$, and gauge properties of the WZNW model.
   
\section{WZNW model on curved manifolds} 
\cleq

Basic properties of two--dimensional WZNW model are defined by the
action 
\be
I(g)=I_0(v) + n\G(v) 
      =\fr{1}{2}\k\int_\S ({}^*v,v) + \fr{1}{3}\k\int_M (v,v^2)\, ,
                               \qquad v\equiv g^{-1}dg\, ,      \leq{2.1}
\ee
where $n$ is  an integer, $\k=n\k_0$, and $\k_0$ is a normalization
constant. The first term is a $\s$--model action which provides
dynamics for a group--valued field $g$, defined over a
two--dimensional, Riemann manifold $\S$, and taking values in a
semi-simple Lie group $G$, while the second term is the topological
Wess--Zumino term, defined on a three--manifold $M$ whose boundary is
$\pd M=\S$. Here, $v$ is the Maurer--Cartan  (Lie algebra valued)
1--form, ${}^*v$ is the dual of $v$, and $(X,Y)=\fr{1}{2}\Tr(XY)$ is the
Cartan--Killing bilinear form on the Lie algebra of G ($\Tr$ denotes
the ordinary matrix trace operation in the adjoint representation of
$G$).  The normalization factor $\k_0$ is chosen in such a way that the
Wess--Zumino term is well defined modulo a multiple of $2\pi$, which is
irrelevant in the functional integral  $Z=\int Dg\exp[iI(g)]$.

Let us now parametrize the group elements by some local coordinates
$q^\a$,  $g=g(q^\a)$, so that
$$
v= E^a t_a \equiv  dq^\a E^a{_\a} t_a \, ,
$$
where $t_a$ are the generators of $G$, satisfying the Lie algebra
$[t_a,t_b]=f_{ab}{^c}t_c$. Then, 
\beqa
&&({}^*v,v)={}^*dq^\a dq^\b\g_{\a\b}\, ,\qquad 
             \g_{\a\b}(q)\equiv E^a{_\a}E^b{_\b}\g_{ab}\, ,\nn \\
&&(v,v^2)=\fr{1}{2}E^aE^bE^cf_{abc}=-6\,d\t \, , \nn
\eeqa
where $\g_{ab}=(t_a,t_b)$ is the Cartan metric on $G$, and
$f_{abc}=f_{ab}{^e}\g_{ec}$. The last equation is based on the 
theorem that any closed form is locally exact. Therefore, the WZNW
action on the group manifold takes the form  
\bsubeq \label{2.2} 
\be
I(q)=\k\int \bigl(\fr{1}{2}{}^*dq^\a dq^\b\g_{\a\b}
                      - dq^\a dq^\b\t_{\a\b} \bigr) \, ,\label{2.2-a}
\ee
where we used $\t= dq^\a dq^\b \fr{1}{2}\t_{\a\b}$. 

Next, we introduce local coordinates $\xi^\m$ $(\m=0,1)$ on $\S$, and
rewrite the action as 
$$
I(q)=\k \int_{\S} d^2\xi \bigl( 
        {\fr 1 2}\sqrt{-g}g^{\m\n}\pd_\m q^\a\pd_\n q^\b \g_{\a\b} 
                  - \ve^{\m\n}\pd_\m q^\a\pd_\n q^\b\t_{\a\b} \bigr)\, ,
$$
where $g^{\m\n}$ is the inverse metric
on $\S$. It is convenient to define an orthonormal basis of tangent
vectors $\pd_i=e_i{^\m}\pd_\m$ $(i=+,-)$, in which the metric  
$\,\eta_{ij}=e_i{^\m}e_j{^\n}g_{\m\n}\,$ takes the light--cone form: 
$\,\eta_{-+}=\eta_{+-}=1$. In this basis,    
\be
I(q)=\k \int_{\S} d^2\xi\sqrt{-g}\,\bigl( 
        {\fr 1 2} \eta^{ij}\pd_iq^\a\pd_j q^\b\g_{\a\b} 
         - \ve^{ij}\pd_iq^\a\pd_jq^\b\t_{\a\b} \bigr)\, . \label{2.2-b}
\ee
\esubeq 
Note that the action $I(q)$ is invariant under conformal
transformations $g_{\m\n}\to g_{\m\n}e^{2F}$  (which implies 
$\pd_i\to e^{-F}\pd_i$). 

We now turn our attention to $G=SL(2,R)$. Using the fact that any 
element g of $SL(2,R)$ in a neighborhood of identity admits the Gauss
decomposition $g=g_+g_0g_-$, where $g_+$, $g_0$ and $g_-$ are defined
in terms of group coordinates $q^\a=(x,\vphi,y)$ as in Eqs.(\ref{B.3}), one
can find explicit expressions for $\g_{\a\b}$ and $\t$, Eqs.\eq{B.5} and
\eq{B.6}, and obtain  
\beqa
I(q) &&= \k \int_{\S}d^2\xi\sqrt{-g}\,\bigl[ 
              {\fr 1 2} \eta^{ij}\pd_i\vphi\partial_j\vphi
        +2(\eta^{ij} -\ve^{ij}) \pd_i x\pd_j y e^{-\vphi} \bigr] \nn \\
        &&= \k \int_{\S}d^2\xi\sqrt{-g}\,\bigl( 
   \pd_+\vphi\pd_-\vphi +4\pd_{+}x\pd_{-}ye^{-\vphi} \bigr)\, .
                                                              \leq{2.3}
\eeqa

\section{Gauge extension of the WZNW action} 
\cleq

We shall now discuss how one can gauge the WZNW theory starting from
the existence of global symmetries, as usual. The action $I(g)$, 
where $g$ belongs to $SL(2,R)$, is invariant under the 
{\it global\/} transformations on the $SL(2,R)$ manifold:   
$$
g\to g'=\O g\bO^{-1}\, ,\qquad dg\to (dg)'=\O(dg)\bO^{-1}\, . 
$$
where $(\O,\bO)$ is an element of $SL(2,R)\times SL(2,R)$.
We want to introduce the corresponding gauge theory, having the
following properties: \par
\bitem
\item[{$a)$}] it should be invariant under the {\it local\/}
transformations   
\be
g'=\O g\bO^{-1} \, , \qquad \O=\O(\xi^-,\xi^+)\, ,
                      \quad \bO=\bO(\xi^+,\xi^-) \, ,      \leq{3.1} 
\ee 
where $(\O,\bO)$ belongs to a subgroup $H$ of 
$SL(2,R)\times SL(2,R)$ (which may be equal to the whole group), \par 
\item[{$b)$}] it should be defined as a field theory on $\S$. \par 
\eitem   
\ni It is well known that the second requirement can not be fulfilled
for every gauge group $H$ \cite{11,12}, since the WZ term $n\G$, originally
defined on $M$, does not have a gauge invariant extension that can be
reduced to an integral over $\S=\pd M$. Possible solutions of this
problem will be discussed after clarifying the meaning of the first
requirement $a)$.

The transformation law of $dg$ under gauge transformation is
changed, but the change can be compensated by introducing the 
{\it covariant derivative\/}:    
\be
Dg\equiv dg+Ag-gB \, ,\qquad (Dg)'=\O(Dg)\bO^{-1} \, ,      \leq{3.2}
\ee
where $(A,B)$ are {\it gauge fields\/} (Lie algebra valued 1--forms).
The covariant derivative $Dg$ transforms homogeneously under local
transformations, provided the gauge fields transform according to  
\be
A'=\O (A + d)\O^{-1} \, ,\qquad B'=\bO (B + d)\bO^{-1} \, . \leq{3.3}
\ee
Having defined $Dg$, one can try to gauge the WZNW action by replacing
$dg\to Dg$, i.e. by replacing 1--form $v=g^{-1}dg$ with the
corresponding covariant 1--form $V$: 
\be
V\equiv g^{-1}Dg=v +g^{-1}Ag -B\, ,\qquad V'=\bO V\bO^{-1}\, .\leq{3.4}
\ee

It is also useful to define the field strengths, $F_A=dA+A^2$ and
$F_B=dB+B^2$, which transform as follows:
$\,F'_A=\O F_A\O^{-1}$, $\,F'_B=\bO F_B\bO^{-1}$.

Now, we apply this procedure to {\it formally\/} define a gauge
invariant extension of the WZNW action \eq{2.1}:
\be
I(g,A,B)=I_0(V)+n\G(V)
    =\fr{1}{2}\k\int_\S ({}^*V,V)+\fr{1}{3}\k\int_M (V,V^2)\, . \leq{3.5}
\ee

The first term $I_0(V)$ is both $a)$ gauge invariant, and $b)$ defined
over $\S$, so that it represents an acceptable gauge invariant action. 
It can be written as
\bsubeq \label{3.6} 
\beqa
&&I_0(V)=I_0(v) + \D_0 \, ,\nn \\
&&\D_0 = \k \int_\S \fr{1}{2}\Tr\bigl[ -{}^*\bv A -{}^*v B
      -{}^*(g^{-1}Ag)B +\fr{1}{2}({}^*AA+{}^*BB)\bigr]\, , \label{3.6-a}
\eeqa
where $\bv=gdg^{-1}=-gvg^{-1}$.

The second term $n\G(V)$ is defined as an integral of a three--form
on $M$, which is gauge invariant. However, this form is in general 
not exact, so that $n\G(V)$ cannot be expressed as an integral
over $\S$; therefore, it can not be used as part of the $\s$--model
action on $\S$.  

We shall now analyze some additional restrictions under which an
acceptable gauge extension of $I(g)$ {\it can\/} be defined. First we
note that, after some algebra, the second term can be rewritten as   
\beqa
&&n\G(V)=n\G(v)+\G_1+\G_2+\G_3 \, ,\nn \\
&&\G_1=\k \int_\S \fr{1}{2}\Tr\bigl[
       -\bv A + vB +g^{-1}AgB \bigr] \, , \nn \\
&&\G_2=\k \int_M\fr{1}{2}\bigl[\cs_3(B,F_B)-\cs_3(A,F_A)\bigr]\, ,\nn \\
&&\G_3=\k \int_M \fr{1}{2}\Tr\bigl[
      F_{A}(Dg)g^{-1} +F_{B}g^{-1}(Dg) \bigr] \, ,        \label{3.6-b}
\eeqa
\esubeq 
where $\cs_3(A,F_A)=\Tr(AF_{A}-\fr{1}{3}A^3)$ is the Chern--Simons
three--form. Formal extension of $I(g)$, obtained in Eqs.\eq{3.5} and
(\ref{3.6}),  can be written as
\beqa
&&I(g,A,B)=I^r(g,A,B)+\G_2(A,B)+\G_3(g,A,B) \, ,\nn\\
&&I^r(g,A,B)\equiv I(g)+\D_0(g,A,B)+\G_1(g,A,B) \, ,       \leq{3.7}
\eeqa
where $\G_2$ and $\G_3$ are defined not on $\S$ but on $M$, violating
thereby the basic requirement $b)$.  
Now, one should observe that the term $\G_3$ is gauge
invariant, therefore it can be removed from $I(g,A,B)$, leaving us with
the gauge invariant combination $I^r(g,A,B)+\G_2(A,B)$. 
Since $\G_2$ is a three--form on $M$, only $I^r(g,A,B)$ can be
included as part of the action for the $\s$--model on $\S$, but it is
not gauge invariant: 
$$
\d I^r(g,A,B) =-\d\G_2(A,B) 
  =-\k \int_M \fr{1}{2}\d\bigl[\cs_3(B,F_B)-\cs_3(A,F_A)\bigr]\ne 0 \, .
$$
The action $I^r(g,A,B)$ is very close to what we want: it is defined on
$\S$, and its variation under gauge transformations gives an expression
which depends on gauge fields $(A,B)$, but not on $g$. Can one find a
mechanism that compensates this non--invariance, and yields an
acceptable gauge invariant extension of the WZNW action \eq{2.1}?   

In the analogous four--dimensional model Witten \cite{13} solved
the problem by requiring the constraint $\cs_3(B,F_B)-\cs_3(A,F_A)=0$ on
the gauge group $H$, sufficient for gauge invariance.
In string models one can simply remove $\G_2$ without
assuming any constraint on $H$, while the gauge invariance of the
theory is ensured by the presence of some additional field in the
action, with ``anomalous" transformation law \cite{12}. In this paper we
shall solve the $\G_2$--problem by considering a gauge extension of the
action \eq{1.1}, describing a system of {\it two\/} simple WZNW models, in
which the problematic $\G_2$ term in the first sector will cancel  the
corresponding term in the second sector, leading to the theory which is
both $a)$ gauge invariant and $b)$ defined on $\S$.  

The construction goes as follows. We start with the formal extension of
$I(g,A,B)$, as obtained in Eq.\eq{3.7}.
Next, using gauge invariance of $\G_3(g,A,B)$ we define a simpler
gauge invariant action:
$$
I'(g,A,B)=I^r(g,A,B) +\G_2(g,A,B)\, .
$$
It is now easy to see that an acceptable gauge extension of the action \eq{1.1}
for the WZNW system can be defined by 
\be
I(1,2)=I'(g_1,A,B)-I'(g_2,A,B) \, ,                          \leq{3.8}
\ee
where $g_1=g(x_1,\vphi_1,y_1)$ and $g_2=g(x_2,\vphi_2,y_2)$ are
different fields, belonging to the same representation of $SL(2,R)$.
Indeed, since $\G_2(A,B)$ does not depend on $g$, the
contribution of two $\G_2$ terms to $I(1,2)$ vanishes. 
Thus, the gauge invariant action \eq{3.8} can be written in the simpler
form   
\bsubeq \label{3.9} 
\be
I(1,2)=I^r(g_1,A,B)-I^r(g_2,A,B) \, ,                    \label{3.9-a}
\ee
where we clearly see that it is an action defined on $\S$. 

The reduced action \eq{3.7} can be written as
\be
I^r(g,A,B) =I(g) +\k \int_\S \fr{1}{2}\Tr\bigl[ 
   -({}^*\bv+\bv)A -({}^*v-v)B -({}^*B +B)(g^{-1}Ag)\bigr]\, ,\label{3.9-b}
\ee
\esubeq 
where the $g$--independent term $\fr{1}{2}({}^*AA+{}^*BB)$ in $\D_0$ is
ignored, as its contribution to sectors 1 and 2 in (\ref{3.9-a}) is
canceled. The absence of this term implies that $I(1,2)$ does not
depend on $({}^*A+A)\sim A_-$ and $({}^*B-B)\sim B_+$ [see Eq.\eq{A.6}] , i.e.
selfdual and anti-selfdual parts of the gauge fields $A$ and $B$,
respectively. As we shall see, the absence of these parts will greatly
influence the dynamical structure of the gauged WZNW system.  

\section{$H_+\times H_-$ gauge theory} 
\cleq

In this section we shall specify the gauge group, and derive an
explicit expression for the gauge action (\ref{3.9}) in terms of the group
coordinates $(x_1,\vphi_1,y_1)$ and $(x_2,\vphi_2,y_2)$.

Using now the matrix $R(g)$, that defines the adjoint representation of
the gauge group,  
$g^{-1}t_bg=-t_cR^c{_b}$,  where $R^c{_b}\equiv E^c{_\a}\bE^\a{_b}$
(Appendix B), the reduced action takes the form 
\be
I^r(g,A,B)=I(g) +2\k \int_\S d^2\xi\sqrt{-g}\,\bigl[  
             -\bv^a_{-}A^b_{+} -v^a_{+}B^b_{-} 
                   +B^a_{-}R^b{_c}A^c_{+}\bigr]\g_{ab}\, .   \leq{4.1}
\ee

As we mentioned, the action $I(1,2)$ does not contain
variables $(A_{-}, B_{+})$, which implies the existence of an 
{\it extra\/} gauge symmetry, allowing an arbitrary change of the
absent components. This is a specific feature of the action for the
WZNW system. To simplify further considerations we shall fix this
symmetry by imposing the following gauge conditions: 
\be
A_{-}=0\, , \qquad B_{+}=0 \, .                              \leq{4.2}
\ee

Up to now we did not specify the gauge group $H$. We could take $H$ to
be the whole $SL(2,R)\times SL(2,R)$, but for our purposes this is not
necessary. We assume that $H$ is a subgroup of $SL(2,R)\times SL(2,R)$, 
defined by 
\be
H= H_+\times H_- \, ,                                        \leq{4.3}
\ee
where $H_+$ and $H_-$ are subgroups of $SL(2,R)$ defined by the generators
$(t_+,t_0)$ and $(t_0,t_-)$, respectively.
When compared to $SL(2,R)\times SL(2,R)$, our choice means that the
gauge fields should be restricted as follows: 
\be
A^\-=0\, ,\qquad  B^\+=0  \, .                              \leq{4.4}
\ee
The gauge symmetry $H_+\times H_-$ is defined in terms of the following
gauge fields and gauge parameters: 
$$
(A^\+_+,A^\0_+, B^\-_-,B^\0_-)\, ,\qquad (\ve^\+,\ve^\0,\bve^\-,\bve^\0)\, .
$$
Gauge transformations of dynamical variables are given in Eqs.(C.4)
and (C.5). 

Using the general relations
$$
v_i =g^{-1}\pd_i g =t_aE^a{_\a}\pd_i q^\a \, ,\qquad
\bv_i= g\pd_i g^{-1}=t_a\bE^a{_\a}\pd_i q^\a \, , 
$$
and assuming the restrictions (4.2) and (4.4), one obtains  
\beqa
\bv^a_{-} A^b_{+}\g_{ab} &&= 2\bv^\-_{-}A^\+_{+} +\bv^\0_{-}A^\0_{+} \nn\\
 &&= -2e^{-\vphi}\pd_- y \bigl[A^\+_{+}+xA^\0_{+}\bigr] 
                                           -\pd_-\vphi A^\0_{+}\, ,\nn\\
v^a_{+} B^b_{-}\g_{ab} &&= 2 v^\+_{+}B^\-_{-} +v^\0_{+}B^\0_{-}  \nn\\
&&= 2e^{-\vphi}\pd_+ x\bigl[B^\-_{-}+yB^\0_{-}\bigr] 
                                           +\pd_+\vphi B^\0_{-}\, .\nn
\eeqa
Next, with the help of the expression (B.8) for $R_{ab}$ we find
\beqa
B^a_{-}R_{ab}A^b_{+}=&&-2e^{-\vphi}\bigl[ yB^\0_{-}A^\+_{+} 
                                    +xy B^\0_{-}A^\0_{+}  \nn\\
      &&\hskip40pt +B^\-_{-}A^\+_{+} +xB^\-_{-}A^\0_{+} \bigr] 
                                     -B^\0_{-}A^\0_{+} \, .\nn
\eeqa
The final result for the reduced action takes the form
\beqa
I^r(g,A,B) = \k\int d^2\xi\sqrt{-g}\,&&\Bigl[ 
\pd_-\vphi\pd_+\vphi+2A^\0_{+}\pd_-\vphi -2B^\0_{-}\pd_+\vphi\nn\\
&&\hskip30pt +4D_+xD_-y\,e^{-\vphi} -2B^\0_{-}A^\0_{+} \Bigr] \, ,\leq{4.5}
\eeqa
where 
$$
D_+ x =\bigl[\pd_+ +A^\0_+\bigr]x +A^\+_+ \, ,\qquad
D_- y =\bigl[\pd_- -B^\0_-\bigr]y -B^\-_-   \, ,
$$
are covariant derivatives on the group manifold, Eqs.(\ref{C.2}).
The last, $g$--independent term in \eq{4.5} will be canceled in the
complete action $I(1,2)$, Eq.(\ref{3.9-a}).  

\section{Induced gravity from gauged WZNW system} 
\cleq

In this section we shall consider the functional integral of the theory
defined by the action (\ref{3.9}), and show, by performing a suitable gauge
fixing and integrating out some dynamical variables, that this theory
leads to the induced gravity action \eq{1.2}. 

\subsection{Effective theory for gauged WZNW system}

It is useful to introduce auxiliary fields $f_{1\pm},f_{2\pm}$, and
rewrite the part of the action $I(1,2)$ given by
\be
\Y\equiv 4D_+x_1D_-y_1 e^{-\vphi_1} -4D_+x_2D_-y_2 e^{-\vphi_2}\, ,
                                                              \leq{5.1}
\ee
in the form
\beqa
\Y =&& f_{1-}D_+x_1 +f_{1+}D_-y_1 -\fr{1}{4}f_{1-}f_{1+}e^{\vphi_1} \nn\\
    && -f_{2-}D_+x_2 -f_{2+}D_-y_2 +\fr{1}{4}f_{1-}f_{1+}e^{\vphi_2} \, ,\nn
\eeqa
or, more explicitly,
\beqa
\Y =&& -B^\-_-(f_{1+}-f_{2+}) +A^\+_+(f_{1-}-f_{2-}) \nn\\
    && +f_{1-}\bigl[\pd_+ +A^\0_+\bigr]x_1 
      +f_{1+}\bigl[\pd_- -B^\0_-\bigr]y_1
                                -\fr{1}{4}f_{1-}f_{1+}e^{\vphi_1} \nn\\
    && -f_{2-}\bigl[\pd_+ +A^\0_+\bigr]x_2 
      -f_{2+}\bigl[\pd_- -B^\0_\-\bigr]y_2
                                +\fr{1}{4}f_{2-}f_{2+}e^{\vphi_2} \, .\nn
\eeqa

Gauge transformations determined by $\ve^\+$ and $\bve^\-$ have the form 
\beqa
&& \d x_1=\ve^\+   \, ,\qquad \d x_2=\ve^\+\, ,
                \qquad \d A^\+_+=-\bigl[\pd_+ +A^\0_+\bigr]\ve^\+ \, ,\nn\\
&& \d y_1=-\bve^\- \, ,\qquad \d y_2=-\bve^\-\, ,
                   \qquad \d B^\-_-=-\bigl[\pd_- -B^\0_-\bigr]\bve^\-\, .\nn
\eeqa
This part of the complete gauge symmetry can be fixed by imposing the
following gauge conditions:  
$$
x_2=0\, , \qquad  y_2=0 \, .                                  
$$

Integrations $\int dA^\+_+dB^\-_-$ and $\int df_{2+}df_{2-}$ in the
functional integral lead to the elimination of $f_{2\pm}$ from the
action: $f_{2\pm}=f_{1\pm}$. Writing $f_{\pm}$ instead of $f_{1\pm}$
for simplicity, one obtains
$$
\Y =  f_{-}\bigl[\pd_+ +A^\0_+\bigr]x_1 
     -f_{+}\bigl[\pd_- -B^\0_-\bigr]y_1
     -\fr{1}{4}f_{-}f_{+}\bigl(e^{\vphi_1} -e^{\vphi_2}\bigr) \, .
$$

After that the integration $\int dx_1dy_1$ produces 
$$
\d \bigl[ (\nabla_+ -A^\0_+)f_{-}\bigr]\cdot
                        \d \bigl[(\nabla_- +B^\0_-)f_{+}\bigr] \, ,
$$
where $\nabla_\pm$ is the covariant derivative on $\S$ (Appendix A).
Then, $\int dA^\0_+dB^\0_-$ yields 
\be
A^\0_+=-\con_+ +\pd_+\ln \vert f_{-}\vert \, ,\qquad 
B^\0_-=-\con_- -\pd_-\ln \vert f_{+}\vert  \, ,               \leq{5.2}
\ee
where $\con_\pm$ is the connection on $\S$ (Appendix A), and an
additional factor $[\det(f_{-}f_{+})]^{-1}$ appears in the functional
measure. Consequently, we find 
\be
\Y =-\fr{1}{4}f_{-}f_{+}\bigl(e^{\vphi_1}-e^{\vphi_2}\bigr)\, ,
                                                             \leq{5.3}
\ee
and the reduced action becomes
\beqa
&&I^r(\vphi, f_-,f_+)=\k\int d^2\xi\sqrt{-g}\,\bigl[ \pd_-\vphi
  \pd_+\vphi-2\bigl(\con_+ -\pd_+\ln \vert f_{-}\vert\bigr)\pd_-\vphi \nn\\
&&\hskip4.5cm  +2\bigl(\con_- +\pd_-\ln \vert f_{+}\vert \bigr)\pd_+\vphi 
   -\fr{1}{4} f_{-}f_{+}e^{\vphi} \bigr] \, .                \leq{5.4}
\eeqa

The complete action $I(1,2)$ is invariant under the remaining
piece of gauge transformations: 
\beqa
&&\d f_{-} =-\ve^\0 f_{-} \, ,\qquad \d f_{+} =\bve^\0 f_{+}\, ,\nn\\ 
&&\d\vphi_1 =\ve^\0 -\bve^\0 \, ,\qquad \d\vphi_2 = \ve^\0-\bve^\0 \, . \nn
\eeqa
Now, we introduce gauge invariant variables,
$$
\phi_1=\vphi_1+\ln \vert f_{+}f_{-}\vert \, ,\qquad 
\phi_2=\vphi_2+\ln \vert f_{+}f_{-}\vert \, ,
$$ 
in terms of which, after some cancellation of $\phi$--independent
pieces, we obtain 
\beqa
&&I(1,2)=I^r(\phi_1)-I^r(\phi_2) \, ,\nn\\
&&I^r(\phi)=\k\int_\S d^2\xi\sqrt{-g}\,\bigl[ 
     \pd_-\phi\pd_+\phi +2\con_-\pd_+\phi -2\con_+\pd_-\phi 
                  -\fr{1}{4} e^\phi \bigr] \, .               \leq{5.5}
\eeqa

Note that the part of the functional integral depending on $f_-,f_+$ is
decoupled from the rest, and can be absorbed into the normalization
factor. Thus, using gauge invariant variables effectively restricts
the space of dynamical variables; it is, essentially, equivalent to a
gauge--fixing corresponding to $(\ve^\0,\bve^\0)$ transformations, e.g. 
$f_{-}=\m_-$, $f_{+}=\m_+$, followed by the integration  
$\int df_{-}df_{+}$. 

Thus, the final form of the effective action for the gauged WZNW system
is given by (5.5). 

\subsection{Transition to the induced gravity}  

To show that the effective theory (5.5) is equivalent to the
induced gravity (1.2), let us observe that (5.5) is invariant under the
following conformal rescalings: 
\be
g_{\m\n}\to e^{2F}g_{\m\n} \, ,\, ,\qquad
\phi_1\to\phi_1-2F \, ,\qquad \phi_2\to \phi_2 -2F \, ,       \leq{5.6}
\ee
which imply $\pd_\pm\to e^{-F}\pd_\pm$, and 
$\con_\pm\to e^{-F}\bigl( \con_\pm \mp \pd_\pm F \bigr)$ . This symmetry is
directly connected to the invariance of the original WZNW theory under
conformal rescalings. It can be gauge--fixed by demanding  
$$
\phi_2=\ln\m \, , 
$$
whereafter the effective action becomes
\be
I(\phi,g_{\m\n})=\k\int_\S d^2\xi\sqrt{-g}\,\bigl[ 
     \pd_-\phi\pd_+\phi +2\con_-\pd_+\phi -2\con_+\pd_-\phi 
            -\fr{1}{4}\m\bigl( e^\phi-1\bigr)\bigr] \, ,       \leq{5.7}
\ee
where we introduced $\phi=\phi_1-\ln\m$.
Now, partial integrations together with equation (A.3), and the
replacement $\phi\to\phi/\sqrt{\k}$, lead to the induced gravity action
(1.2), where $\a=2\sqrt{\k}$, and $M=\k\m/4$.

\section{Concluding remarks} 
\cleq

We presented here the connection between the gauge extension of the
WZNW system (1.1) and the induced gravity action (1.2), fully
respecting the diffeomorphism invariance of both theories. 

It is well known that an acceptable gauge extension of the simple WZNW
model does not exist unless one requires specific constraints on the 
gauge group \cite{11}. The reason for this unusual behaviour steams from the
fact that gauged WZ term $n\G$ does not represent, in general, a
field theory on a 2D manifold $\S$. If one tries to select an action
defined on $\S$, one looses gauge invariance, and vice versa.  
In string models one can overcome these problems with the help of
an additional field \cite{12}. Following the ideas developed in 
Ref. \cite{10}, we introduced in this paper an acceptable action, which
is both gauge invariant and defined on $\S$, by considering a dynamical
system described by a difference of two simple WZNW models. 

Our gauge group is $H_+\times H_-$, a four--parameter subgroup of
$SL(2,R)\times SL(2,R)$. Two of the gauge fields, $A^\+_+$ and
$B^\-_-$, ensure the equality of currents in two sectors, 
in accordance with the results of the Hamiltonian analysis of the WZNW
system \cite{10}. The remaining two fields, $A^\0_+$ and $B^\0_-$, become
components of the connection of the induced gravity action. In this
way, the Riemannian structure on $\S$ is seen to be closely related
to the $SL(2,R)$ gauge fields.

The results obtained here can be used to clarify the connection between
globally regular solutions of the equations of motion for the WZNW
system, and the related singular solutions (in the coordinate sense) 
of the induced gravity \cite{8,9}. In particular, it will be interesting to
improve our understanding of the WZNW black hole solutions in the
context of induced gravity \cite{14}. 

\appendix 

\section{Riemannian structure of $\S$} 

In this Appendix we present some formulae on the Riemannian structure
of 2D manifold $\S$, which are used in the paper.

Coordinates of points in two--dimensional manifold $\S$ are denoted by
$\xi^\m$ $(\m=0,1)$. Basic tensorial objects in the coordinate basis are:
\bitem
\item[{}] vectors: ${e}_\m=\pd_\m\,$;~~~
       1--forms: $\th^\m=d\xi^\m\,$;~~~
       metric: ${e}_\m\cdt{e}_\n=g_{\m\n}\,$;~~~$\ve^{01}=1\,$; 
\eitem
\ni Another useful basis is the local light--cone basis: 
\bitem
\item[{}] vectors:   ${e}_i =\pd_i=e_i{^\m}{e}_\m\,$;~~~
  1--forms: $\th^i=d\xi^i =e^i{_\m}d\xi^\m\,$~~$(i=+,-)\,$; 
\item[{}] metric: ${e}_i\cdt{e}_j=e_i{^\m}e_j{^\n}g_{\m\n}=\eta_{ij}$,~~
                $\eta_{+-}=\eta_{-+}=1\,$;~~~ $\ve^{-+}=1\,$;
\eitem  
\subsub{Connection and curvature.} Riemannian connection on $\S$,
$\con^i{_j}=\ve^i{_j}\con$, is defined by the first structural equation:  
\be
d\th^i+\con^i{_j}\th^j=0\, .                                    \leq{A.1}
\ee 

The exterior derivative of a 1--form $u=u_k\th^k$ can be written as
$$
d u=(d u_i)\th^i +u_id\th^i =(d u_i -u_s\con^s{_i})\th^i 
       =(\nabla_ku_i)\th^k\th^i \, ,
$$
where $\nabla_ku_i$ is the covariant derivative of a 1--form:
$$
\nabla_k u_i  = \pd_ku_i-\ve^s{_i}\con_k u_s \, ,\qquad
\nabla_k u_\mp = (\pd_k \mp\con_k)u_\mp  \, .
$$
By noting that $u^\pm=u_\mp$, one easily finds the covariant derivative
of a vector.  

The curvature is defined by the second structural equation: 
\be
d\con^i{_j}=\fr{1}{2}R^i{_{jkl}}\th^k\th^l \, .                  \leq{A.2} 
\ee
Using $d\con=(\nabla_k\con_l)\th^k\th^l$, one finds
\beqa
&&R^i{_{jkl}}=\ve^i{_j}\bigl(\nabla_k\con_l-\nabla_l\con_k\bigr)\, ,\nn\\
&&R=2R_{-+} =2\bigl(\nabla_-\con_+ -\nabla_+\con_- \bigr)\, .   \leq{A.3}
\eeqa
\subsub{Conformal rescaling.} Let us now derive the transformation law
of the connection under conformal rescaling of the metric.
The relation $g_{\m\n}=e^{2F}\hg_{\m\n}$ implies $\th^i=e^F\hth^i$.
Replacing this into the first structural equation gives  
$$
d\hth^i +dF\hth^i +\con^i{_k}\hth^k =0 \quad \Ra \quad
\hcon^i{_j}\hth^j = dF\hth^i +\con^i{_k}\hth^k \, .
$$
Since $\hcon^i{_j}=\ve^i{_j}\hcon$, we easily obtain
\be
\hcon_+ -\hat\pd_+F =\con_+ e^F \, , \qquad 
\hcon_- +\hat\pd_-F =\con_- e^F \, .                           \leq{A.4}
\ee
Acting on these equations with $\hnabla_-$ and $\hnabla_+$,
respectively, and using (A.4) again, one finds
\beqa
&&\hnabla_-\hcon_+ -\hnabla_-\hpd_+F=e^{2F}\nabla_-\con_+ \, ,\nn\\
&&\hnabla_+\hcon_- +\hnabla_+\hpd_-F=e^{2F}\nabla_+\con_- \, . \nn
\eeqa
Subtracting two equations one finds the effect of conformal rescaling
on the curvature:
\be
R(\hg)-2{\eta^{ij}{\hat\nabla}_i{\hat\nabla}_j}F=e^{2F}R(g)\, .\leq{A.5}
\ee

We display here some useful formulae:
\beqa
&&\th^-\th^+=d^2\xi\sqrt{-g}\, ,\qquad {}^*\th^i=\ve^i{_k}\th^k\, ,\nn\\
&&\th^i\th^j=\ve^{ij}\, d^2\xi\sqrt{-g} \, ,\qquad 
            {}^*\th^i\th^j=\eta^{ij}\, d^2\xi\sqrt{-g} \, ,\nn\\
&&A\pm{}^*A=2A_{\mp}\th^{\mp} \, ,\qquad
       ({}^*A \pm A)B= 2A_\mp B_\pm\, d^2\xi\sqrt{-g} \, .    \leq{A.6} 
\eeqa

\section{On the geometry of $SL(2,R)$ manifold} 
\cleq

Here we present some useful results concerning the Riemannian
structure of the group manifold $SL(2,R)$.  

If the generators of the group $SL(2,R)$ are chosen as
$t_{(\pm)}={\fr 1 2}(\s_1\pm i\s_2)$, $t_\0 ={\fr 1 2}\s_3$, where
$\s_k$ are the Pauli matrices, the Lie algebra
$[t_a,t_b]=f_{ab}{^c}t_c$ takes the form 
\be
[t_\+,t_\-]=2 t_\0 \, ,\qquad [t_{(\pm)},t_\0]=\mp t_{(\pm)}\, .
                                                               \leq{B.1}
\ee
Then, the calculation of the Cartan metric 
$\g_{ab}=(t_a,t_b)={\fr 1 2}f_{ac}{^d}f_{bd}{^c}$ yields 
\be
\g_{ab}=\pmatrix{ 0 &0 &2\cr
                  0 &1 &0\cr
                  2 &0 &0\cr} \, ,\qquad a,b={\+,\0,\-}\, .    \leq{B.2}
\ee
Raising and lowering of the tangent space indices $(a,b,...)$ are
performed with $\g_{ab}$ and its inverse $\g^{ab}$. 

The group $SL(2,R)$ has the property that any element $g$ in a
neighborhood of identity admits the Gauss decomposition: 
\bsubeq \label{B.3} 
\beqa
&&g=g_+(x)g_0(\vphi)g_-(y) \, ,\nn\\
&&g_+=e^{xt_\+}=1+x t_\+\, , \qquad  g_-=e^{yt_\-}=1+yt_\- \, ,\nn\\
&&g_0=e^{\vphi t_\0}= \cos(\vphi/2)+2t_\0\sin(\vphi/2) \, .  \label{B.3-a}
\eeqa 
where $q^\a=(x,\vphi,y)$ are group coordinates. In this parametrization
we have 
\be
g=\pmatrix{ e^{\vphi/2}+xye^{-\vphi/2} &xe^{-\vphi/2} \cr
          ye^{-\vphi/2}              &e^{-\vphi/2} \cr }\, . \label{B.3-b}
\ee
\esubeq 

Now, we can write $v=g^{-1}dg=E^at_a=t_a E^a{_\a}dq^\a$, where the
quantities $E^a{_\a}$ serve as the vielbein on the group manifold.
The above expression for $g$ leads to
\beqa
&&E^\+=e^{-\vphi}dx \, ,\nn\\
&&E^\0=2ye^{-\vphi}dx +d\vphi \, ,\nn\\
&&E^\-=-y^2e^{-\vphi}dx -yd\vphi +dy \, , \nn
\eeqa
so that the vielbein $E^a{_\a}$ and its inverse $E^\a{_a}$ are given as
\be
E^a{_\a} = \pmatrix{   e^{-\vphi} &    0&   0\cr
                      2ye^{-\vphi}&    1&   0\cr
                    -y^2e^{-\vphi}&   -y&   1\cr} \, ,\qquad
E^\a{_a} = \pmatrix{   e^\vphi &    0&     0\cr
                        -2y    &    1&     0\cr
                        -y^2   &    y&     1\cr} \, .           \leq{B.4}
\ee
The Cartan metric in the coordinate basis, 
$\g_{\a\b}=E^a{_\a}E^b{_\b}\g_{ab}$, has the form 
\be
\g_{\a\b}=\pmatrix{    0      &  0& 2e^{-\vphi}\cr
                       0      &  1&      0\cr 
                   2e^{-\vphi}&  0&      0\cr }  \, ,
                                 \qquad \a,\b=x,\vphi,y\, .     \leq{B.5}
\ee

From the relation $(v,v^2)=-6d\t$ we obtain 
\be
d\t=E^\+ E^\0 E^\- =d\bigl( e^{-\vphi}dx dy \bigr)\, .          \leq{B.6}
\ee

Similarly, the calculation of $\bv=gdg^{-1}=t_a\bE^a=t_a\bE^a{_\a}dq^\a$ 
leads to
\beqa
&&\bE^\+=-dx + xd\vphi + x^2e^{-\vphi}dy \, ,\nn\\
&&\bE^\0=\hskip31pt -d\vphi  -2xe^{-\vphi}dy\, ,\nn\\
&&\bE^\-=\hskip64pt -e^{-\vphi}dy \, , \nn 
\eeqa
or
\be
\bE^a{_\a}=\pmatrix{-1 &    x &   x^2e^{-\vphi}\cr
                        0 &   -1 &   -2xe^{-\vphi}\cr
                        0 &    0 &   -e^{-\vphi}  \cr} \, ,\qquad
\bE^\a{_a}=\pmatrix{ -1 &   -x &      x^2\cr
                      0 &   -1 &      2x \cr
                      0 &    0 &   -e^{\vphi}  \cr} \, .         \leq{B.7}
\ee
The metric $\bar\g_{\a\b}$ is the same as $\g_{\a\b}$.

Also, we shall be making use of the matrix $R^a{_b}$,  
defined by $g^{-1}t_bg =-t_aR^a{_b}$. Starting from the 
identity  $g^{-1}\bv g=-v$, which can be written in the form 
$\bE^a{_\a}g^{-1}t_ag=-E^a{_\a}t_a$, one finds 
$R^a{_b}=E^a{_\a}\bE^\a{_b}$.
The calculation of $R_{ab}=\g_{ac}R^c{_b}$ yields
\beqa
R_{ab}(g)= \pmatrix{ 2y^2e^{-\vphi}& 2xy^2e^{-\vphi}+2y &
                                -2x^2y^2e^{-\vphi}-4xy-2e^{\vphi} \cr       
               -2ye^{-\vphi}& -2xye^{-\vphi}-1& 2x^2ye^{-\vphi}+2x \cr 
                -2e^{-\vphi}&  -2xe^{-\vphi}  & 2x^2e^{-\vphi}\cr }\, .
                                                              \leq{B.8}
\eeqa

\section{Covariant derivative and gauge transformations} 
\cleq

In this Appendix we exhibit gauge properties of the WZNW system in some
detail. 
\subsub{1.} Writing the expression (3.4) for the covariant 1--form $V$
in group coordinates $q^\a$, one can obtain coordinate expression for the
covariant derivative $Dq^\a$ on the group manifold:  
\beqa
&&t_aE^a{_\a}Dq^\a =t_aE^a{_\a}dq^\a -t_a R^a{_b} A^b -t_aB^a \, ,\nn\\
&&Dq^\a =dq^\a -\bE^\a{_a}A^a - E^\a{_a}B^a \, .               \leq{C.1}
\eeqa
Effectively, the components $A_-$ and $B_+$ are absent, Eq.(4.2), so that  
$$
D_+q^\a =\pd_+q^\a - \bE^\a{_a} A^a_+\, ,\qquad
D_-q^\a =\pd_-q^\a - E^\a{_b}B^b_- \, .
$$
Taking into account additional conditions $A^-_+=B^+_-=0$, Eq.(4.4),
one finds
\bsubeq \label{C.2} 
\beqa
&&D_+ x =\bigl[\pd_+ +A^\0_+\bigr]x +A^\+_+ \, ,\nn\\
&&D_+\vphi =\pd_+\vphi +A^\0_+ \, ,\nn\\
&&D_+ y =\pd_+ y \, ,                                      \label{C.2-a} 
\eeqa
and
\beqa
&&D_- x =\pd_- x  \, ,\nn\\
&&D_-\vphi =\pd_-\vphi -B^\0_- \, ,\nn\\
&&D_- y =\bigl[\pd_- -B^\0_-\bigr]y  -B^\-_-   \, .         \label{C.2-b} 
\eeqa
\esubeq 
\subsub{2.} Let us now consider $SL(2,R)\times SL(2,R)$ gauge
transformations. Group elements transform according to
$g'=\O g\bO^{-1}$, where $\O=e^\ve$, $\bO=e^\bve$, and $\ve=t_a\ve^a$,
$\bve=t_a\bve^a$. Infinitesimal transformations are 
\beqa
&&g^{-1}\d g=g^{-1}\ve g-\bve \, ,\nn\\
&&\d q^\a =-\bE^\a{_a}\ve^a -E^\a{_a}\bve^a \, ,                 \leq{C.3}  
\eeqa
or, in components:
\beqa
&&\d x=\ve^\+ +x\ve^\0 -x^2\ve^\- -e^\vphi\bve^\+ \, ,\nn\\
&&\d\vphi=\hskip35pt \ve^\0 -2x\ve^\- +2y\bve^\+ -\bve^\0 \, ,\nn\\
&&\d y= \hskip66pt e^\vphi\ve^\- +y^2\bve^\+ -y\bve^\0 -\bve^\- \, .\nn
\eeqa

Infinitesimal transformations of gauge potentials are obtained from \eq{3.3}:
$$
\d A^a=-d\ve^a-f_{bc}{^a}A^b\ve^c \, ,\qquad
\d B^a=-d\bve^a-f_{bc}{^a}B^b\bve^c \, .                          
$$
Gauge fixing \eq{4.2} leads to
$$
\d A^a_+= -\pd_+\ve^a -f_{bc}{^a}A^b_+\ve^c \, ,\qquad
\d B^a_-= -\pd_-\bve^a -f_{bc}{^a}B^b_-\bve^c \, .                
$$

\subsub{3.} Now, restriction to $H_+\times H_-$ is achieved by
demanding $\ve^\-=\bve^\+=0$. The restricted transformations take the 
form:
\beqa
&&\d x=\ve^\+ +x\ve^\0 \, ,\nn\\
&&\d\vphi= \ve^\0 -\bve^\0 \, ,\nn\\
&&\d y=  -y\bve^\0 -\bve^\- \, ,                                 \leq{C.4}
\eeqa
and
\beqa
&&\d A^\0_+= -\pd_+\ve^\0 \, ,\qquad
 \d A^\+_+= -[\pd_+ +A^\0_+]\ve^\+ +A^\+_+\ve^\0 \, ,\nn\\
&&\d B^\0_-= -\pd_-\bve^\0 \, ,\qquad
 \d B^\-_-= -[\pd_- -B^\0_-]\bve^\- -B^\-_-\bve^\0 \, .            \leq{C.5}
\eeqa

\subsub{4.} Using $\d V^a= f_{bc}{^a}\bve^b V^c$, where
$V^a=E^a{_\a}Dq^\a$, one finds 
$$
\d(Dq^\a)=\bigl[(E^\a{_a}f_{bc}{^a}E^c{_\b})\bve^b 
                      -E^\a{_a}(\d E^a{_\b})\bigr]Dq^\b \, .
$$
Restriction to $H_+\times H_-$  $[\ve^\+ =\bve^\- =0]$ yields
$$
\d(Dx)=\ve^\0 Dx\, ,\qquad\d(D\vphi)=0\, ,\qquad\d(Dy)=-\bve^\0 Dy\, .
$$
Then, gauge transformations of auxiliary fields $f_{\pm}$ are:
$\,\d f_{+}=\bve^\0 f_{+}$, $\,\d f_{-}= -\ve^\0 f_{-}$. 
From this it follows $\d (\vphi +\ln\vert f_{-}f_{+}\vert ) =0$.

\end{document}